\documentclass[aps,prd,notitlepage,showpacs,nofootinbib,superscriptaddress]{revtex4-1}

\usepackage{graphicx}
\usepackage[utf8]{inputenc} 

\usepackage{amsmath}
\usepackage{amssymb}
\usepackage{float}
\usepackage{comment}
\usepackage{slashed}
\usepackage[normalem]{ulem}
\usepackage{xfrac}
\usepackage{bm}

\usepackage{booktabs}

\usepackage{mathtools,braket,blkarray}

\usepackage{amsmath}
\usepackage{amssymb}

\usepackage[usenames,dvipsnames]{color}
\usepackage[colorinlistoftodos]{todonotes}
\usepackage[colorlinks=true,citecolor=darkred,urlcolor=darkred, pdfborder={0 0 0}]{hyperref}
\usepackage[normalem]{ulem}

\usepackage{multirow}
\definecolor{darkred}{rgb}{0.6,0,0}
\definecolor{darkgreen}{rgb}{0.0, 0.4, 0.0}
\usepackage[colorinlistoftodos]{todonotes}

\definecolor{linkcolor}{rgb}{0,0,0.5}

\definecolor{brown}{rgb}{0.59, 0.29, 0.0}

\newcommand {\ignore}[1]{}

\def \znbb {$\rm 0\nu\beta\beta$ }

\def\gsim{\raise0.3ex\hbox{$\;>$\kern-0.75em\raise-1.1ex\hbox{$\sim\;$}}}
\def\lsim{\raise0.3ex\hbox{$\;<$\kern-0.75em\raise-1.1ex\hbox{$\sim\;$}}}

\def\SM{$\mathrm{SU(3)_c \otimes SU(2)_L \otimes U(1)_Y}$ }
\newcommand{\sm}{{Standard Model }}

\usepackage{soul}

\definecolor{mightnightblue}{RGB}{25,25,112}

\definecolor{brown}{rgb}{0.59, 0.29, 0.0}

\def\vev#1{\left\langle #1\right\rangle}
\def\SM{$\mathrm{SU(3)_c \otimes SU(2)_L \otimes U(1)_Y}$ }
\def\21{$\mathrm{SU(2)_L \otimes U(1)_Y}$}

\def\sm{standard model }

\newcommand{\AddrAHEP}{%
  AHEP Group, Institut de F\'{i}sica Corpuscular --
  C.S.I.C./Universitat de Val\`{e}ncia, Parc Cient\'ific de Paterna.\\
 C/ Catedr\'atico Jos\'e Beltr\'an, 2 E-46980 Paterna (Valencia) - SPAIN}
\newcommand{\AddrDCP}{%
  Dual CP Institute of High Energy Physics, C.P. 28045, Colima, M\'exico}

\begin{document}


\title{\boldmath \color{BrickRed} Scotogenic majorana neutrino masses in a predictive orbifold theory of flavour}

\author{Francisco J. de Anda}\email{fran@tepaits.mx}
\affiliation{Tepatitl{\'a}n's Institute for Theoretical Studies, C.P. 47600, Jalisco, M{\'e}xico}
\affiliation{\AddrDCP}
\author{Omar Medina}\email{Omar.Medina@ific.uv.es}
\affiliation{\AddrAHEP}
\author{Jos\'{e} W. F. Valle}\email{valle@ific.uv.es}
\affiliation{\AddrAHEP}
\author{Carlos A. Vaquera-Araujo}\email{vaquera@fisica.ugto.mx}
\affiliation{Consejo Nacional de Ciencia y Tecnolog\'ia, Av. Insurgentes Sur 1582. Colonia Cr\'edito Constructor, Del. Benito Ju\'arez, C.P. 03940, Ciudad de M\'exico, M\'exico}
\affiliation{Departamento de F\'isica, DCI, Campus Le\'on, Universidad de
Guanajuato, Loma del Bosque 103, Lomas del Campestre C.P. 37150, Le\'on, Guanajuato, M\'exico}
\affiliation{\AddrDCP}

\begin{abstract}
\vspace{0.5cm}

The use of extra space-time dimensions provides a promising approach to the flavour problem.
The chosen compactification of a 6-dimensional orbifold implies a remnant family symmetry $A_4$. 
This makes interesting predictions for quark and lepton masses, for neutrino oscillations and neutrinoless double beta decay,
providing also a very good global description of all flavour observables.
Due to an auxiliary $\mathbb{Z}_4$ symmetry, we implement a scotogenic Majorana neutrino mass generation mechanism with a viable WIMP dark matter candidate.

\end{abstract}

\maketitle
\noindent

\section{Introduction}
\label{Sect:intro}

The \sm lacks a basic principle that one may use to describe flavour properties. 
Indeed, the ``flavour problem'' constitutes one of the major challenges in our field.
Describing the observed pattern of fermion masses and mixing has become even a tougher problem,
after the discovery of neutrino oscillations~\cite{McDonald:2016ixn,Kajita:2016cak,deSalas:2020pgw,zenodo}.
Oscillations demonstrate that leptons do mix, but rather differently from the way quarks do within the Cabibbo–Kobayashi–Maskawa (CKM) model.
Moreover, it seems unlikely that the fermion mass pattern results just by chance.
It seems to suggest the existence of a ``family'' symmetry, for which mathematics offers us many possibilities~\cite{Ishimori:2010au}. 
Extra space-time dimensions may help shed fresh light on the flavour problem.
Indeed, the fermion mass hierarchies may result from geometry~\cite{Arkani-Hamed:1999ylh}, 
while the mixing angles may be related as a consequence of the imposition of suitable symmetries~\cite{Chen:2015jta,Chen:2020udk}.\\[-.4cm] 

Here we build upon our previous proposal of using a class of six-dimensional orbifold constructions as theories of flavour~\cite{deAnda:2019jxw,deAnda:2020pti}. 
In sharp contrast with the warped flavordynamics scenario proposed in~\cite{Chen:2015jta,Chen:2020udk}, where the family symmetries were postulated \emph{ab initio},
now the family symmetry is dictated by the compactification to be $A_4$, 
the smallest group with three-dimensional irreducible representations that we can use to stack the three observed particle families. 

Another important drawback of the \sm is the absence of a viable dark matter candidate.
In our current flavour model construction, dark matter can be identified as the mediator of neutrino mass generation.
Indeed, with the help of an auxiliary symmetry, we implement a scotogenic picture where neutrinos acquire a Majorana mass through the exchange of a dark sector.
The model naturally predicts the ``golden'' quark-lepton mass relation, also making predictions for neutrino oscillations and neutrinoless double beta decay.
Moreover, it provides an excellent global description of all flavour observables. \\[-.4cm]

  The paper is organised as follows. In section~\ref{Sec:TheoreticalFrame} we discuss general features of the theoretical framework.
In Sec.~\ref{sec:TheFlavorModel} we introduce the new fields and their interactions, describing both the scalar as well as the fermionic sector.
In Sec.~\ref{sec:fermion-masses} we discuss fermion masses, including the scotogenic neutrino mass generation mechanism.
 Section~\ref{sec:analysis} contains a discussion on the main flavour predictions of our model.  Finally, we give our conclusions in section \ref{sec:Conclusions}.

\section{Theoretical preliminaries}
\label{Sec:TheoreticalFrame}

Here we consider a theoretical framework with new dimensions \cite{Altarelli:2005yp,Altarelli:2006kg,Adulpravitchai:2009id,Adulpravitchai:2010na,Altarelli:2010gt}
 as a setup to tackle the family problem.
As an alternative to the warped flavordynamics scenario proposed in~\cite{Chen:2015jta,Chen:2020udk} in this paper we study a realistic theory of flavour based 
on a discrete $A_4$ family symmetry that emerges naturally from the orbifold compactification of a 6-dimensional quantum field theory.
This builds upon the proposal made in~\cite{deAnda:2019jxw}.
Here we summarize the main features of the extra-dimensional construction. 

In the 6-dimensional theory, the spacetime manifold $\mathcal{M}$ is identified with the direct product $\mathcal{M} = \mathbb{M}^4\times (T^2/\mathbb{Z}_2)$, 
where $\mathbb{M}^4$ is  the four-dimensional Minkowski spacetime, and $T^2 / \mathbb{Z}_2$ is a one-parameter ($\theta$) family of 2D toroidal orbifolds defined 
by the following relations satisfied by the extra-dimensional coordinates: 
 \begin{align}
 \left(x^5,x^6\right) &= \left( x^5 + 2 \pi R_1, x^6\right), \label{eq:id1} \\
 \left(x^5,x^6\right) &= \left( x^5 + 2 \pi R_2 \cos \theta,  x^6 + 2 \pi R_2 \sin \theta\right) ,\label{eq:id2} \\
  \left(x^5,x^6\right)&= \left(-x^5,-x^6\right) ,\label{eq:id3}
 \end{align}
where the first two equations define a torus, with $\theta$ parameterizing its twist angle, and the third equation defines the $\mathbb{Z}_2$ orbifolding. 
For concreteness, here we assume that the characteristic radii of the compact extra dimensions are of the same order of magnitude, satisfying
 \begin{equation}
 R_1 \sim R_2 \sim 1/M_c,
 \label{eq:cScale}
\end{equation}
in terms of the compactification scale $M_c$. Moreover, we assume that the twist angle takes the value $\theta = 2\pi /3$. 
To simplify the analysis it is convenient to define the scaled complex coordinate $z= M_c(x_5+ i x_6)/(2\pi) $. Hence we can write Eqs. (\ref{eq:id1})-(\ref{eq:id3}) as
 \begin{align}
z &= z+1, \label{eq:id1c} \\
z &= z+\omega,\label{eq:id2c} \\
z&= -z, \label{eq:id3c}
 \end{align}
where $\omega$ is the cubic root of unity 
\begin{equation}
\omega \equiv e^{i \theta}  = e^{i 2\pi / 3} . 
\label{eq:omega}
\end{equation}
\begin{figure}
\includegraphics[scale=0.7]{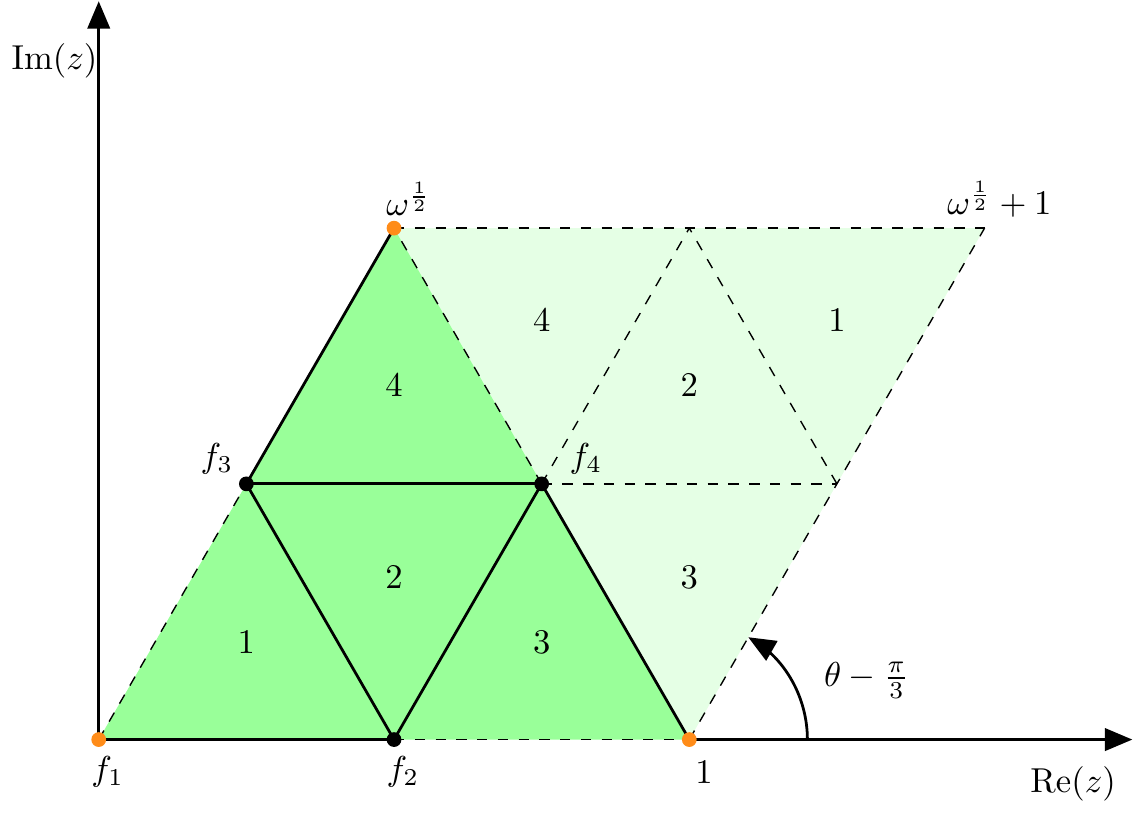}
\caption{ 
  The fundamental domain of the $T^2/\mathbb{Z}_2$ orbifold is shown in the darkest shade of green, obtained after the compactification of the corresponding domain of the twisted torus,
  which includes the region depicted in the lightest shade of green. The resulting space is reminiscent of a tetrahedron, and can be visualized by identifying the three orange dots into a
  single vertex. The fixed points of the orbifold are located at the vertices of the tetrahedron. }
\label{fig:OrbifoldConstruction}
\end{figure}

A fundamental property of orbifolds is that they have singular points. 
In our $T^2/\mathbb{Z}_2$ orbifold, these are located at the points that remain fixed after performing the transformations listed in Eqs.~(\ref{eq:id1c})-(\ref{eq:id3c}). 
Notice that there are four fixed points in our construction
 \begin{equation}
f_1=0,\quad f_2=\frac{1}{2},\quad f_3=\frac{\omega^{\frac{1}{2}}}{2},\quad f_4=\frac{1+\omega^{\frac{1}{2}}}{2}.
 \label{eq:fixed}
\end{equation}
These fixed points define the location of 4-dimensional branes embedded in the 6-dimensional space $\mathcal{M}$.
In Figure \ref{fig:OrbifoldConstruction} both the fundamental domain of the twisted torus $T^2$ (light shaded green),
and the fundamental domain of the $T^2/\mathbb{Z}_2$ orbifold (dark shaded green) are displayed.
After compactification, the continuous Poincar\'e symmetry of the two extra dimensions is broken, leaving a residual $A_4$ symmetry of the branes \cite{Altarelli:2006kg,Burrows:2009pi,Altarelli:2010gt}.
The emergence of the discrete $A_4$ symmetry can be understood from the invariance under permutations displayed by the four fixed points of the orbifold.
Any of these permutations can be written in terms of two independent transformations
\begin{equation}
 S: z\longrightarrow z+1/2, \quad T:z \longrightarrow \omega^2 z.
 \label{eq:permut1}
 \end{equation}
These transformations can be also written as elements of the permutation group $S_4$.
  \begin{equation}
 S=(12)(34), \quad T=(1)(243),
  \label{eq:permut2}
 \end{equation}
In this form, $S$ and $T$ are related to the generators of the $A_4$ group, satisfying its presentation equation:
\begin{equation}
S^2 =T^3=(ST)^3=1.
\label{eq:presentation}
\end{equation}

The model, introduced in the next section, is based on this remnant $A_4$ as a family symmetry \cite{Ma:2001dn,Babu:2002dz,King:2013hj,Morisi:2013eca}.
Equally charged fields located on the four different branes would transform into each other by the remnant $A_4$ transformations.
The four branes transform as the reducible representation $\textbf{4}$, which in turn can be written as a sum of irreducible representations $\textbf{4}\to \textbf{3}+\textbf{1}$.
Thus, the fields localized on the branes must transform under the flavour group $A_4$ as triplets or singlets. The family symmetry $A_4$ is then spontaneously broken,
giving rise to the mass differences between the three generations of fermions, and consequently, quark and lepton mixing. Models based on a similar geometrical construction
have proven to be highly predictive and minimalist \cite{deAnda:2019jxw,deAnda:2020pti,deAnda:2020ssl}.

Notice that the $A_4$ symmetry arises naturally on the branes, so localized fields automatically preserve it.
  The fields that propagate in the bulk preserve the full 6-dimensional Poincar\'e  symmetry, which has $A_4$ as a subgroup.
  When adding fields in the bulk one must assign their transformation properties under $A_4$ in the reduced 4-D theory.
  In our model we have singlet fields $\psi$ on the bulk which are required to satisfy
\begin{equation}
\begin{array}{lll}
1:& T\psi(x,z)=\psi(x,\omega^2 z)=\psi(x, z), & S\psi(x,z)=\psi(x,z+1/2)=\psi(x,z) \\
1':& T\psi(x,z)=\psi(x,\omega^2 z)=\omega \psi(x, z), & S\psi(x,z)=\psi(x,z+1/2)=\psi(x,z) \\
1'':& T\psi(x,z)=\psi(x,\omega^2 z)=\omega^2 \psi(x, z), & S\psi(x,z)=\psi(x,z+1/2)=\psi(x,z).\\
\end{array}
\end{equation}

Note that in the symmetries of Eq. \ref{eq:permut1}, we could also have listed $U:z\longrightarrow z^*$, which is also a symmetry of the fixed points.
  This would enlarge the remnant symmetry to $S_4$ \cite{Ma:2001dn,Babu:2002dz,King:2013hj,Morisi:2013eca}.
  The chosen representations previously described would not preserve this symmetry and only $A_4$ would remain as desired. 

\section{The flavour model} 
\label{sec:TheFlavorModel}

The model is a 6-dimensional extension of the standard model, based on the \SM gauge symmetry, featuring the orbifold compactification described in the previous section.

It inherits a natural $A_4$ discrete family symmetry. The field content and the transformation properties under the different symmetry groups are shown in Table \ref{tab:FieldContent}. 
Note that all fermionic fields, except for the right-handed quarks, transform as flavour triplets and are localized into the orbifold branes.
The decomposition of the 6-dimensional fields into 4-dimensional representations is shown in Appendix \ref{app:6decomp}.

It is important to note that the 6-dimensional chiral fermions $\textbf{u}^c_i$ would generate a 6-dimensional gauge and gravitational anomaly.
  However, the present model is free from 4-dimensional anomalies, and thus, those 6-dimensional anomalies do not affect the phenomenology of the low-energy effective theory of the zero modes.
 We expect our model to be part of a larger UV-complete construction. In our bottom-up spirit, focused on the low-energy phenomenology, such completion lies beyond the scope of the paper.

The scalar sector consists of three weak iso-doublets, two of these are flavour triplets $H_u, H_d$, and the other is a flavour singlet $\eta$.
Moreover, the model includes an SM singlet scalar $\sigma$ that drives the spontaneous breaking of both family symmetry and lepton number,
thus playing the role of a flavoured Majoron~\cite{Chikashige:1980ui,Schechter:1981cv}.

The model assumes an auxiliary $\mathbb{Z}_4$ symmetry which dictates how the weak iso-doublets $H_u, H_d$ couple with fermions,
and forbids neutrino masses at tree level \cite{Chulia:2016ngi,Srivastava:2017sno}.
This $\mathbb{Z}_4$ symmetry is spontaneously broken as $\mathbb{Z}_4\to\mathbb{Z}_2$, where the preserved $\mathbb{Z}_2$ symmetry defines the ``dark sector", 
as it forbids the decay of the lightest $\mathbb{Z}_2$ charged field, providing the stabilizing mechanism needed to describe a viable WIMP dark matter candidate. 
This dark sector includes the inert iso-doublet ($\eta$) that transforms as an $A_4$ singlet, as well as a ``dark fermion" triplet $F$. 
\begin{table}[H] 
\centering
\begin{tabular}{@{}|c|ccc|c|c|c|c@{}}
 \multicolumn{7}{c}{\textbf{Field Content and Quantum Numbers}}                 \vspace{2pt} \\ \hline 
Field &                 $\mathrm{SU(3)_{C}}$       & $\mathrm{SU(2)_{L}}$ & $\mathrm{U(1)_{Y}}$ & $\mathbb{Z}_{4}$ & $A_4$ & Location  \\ \hline
   $L$                   &        $\bm{1}$     &   $\bm{2}$   &   $-1$               &       $1$    &   $\bm{3}$          &   Brane          \\
   $d^{c}$             &        $\bm{3}$     &   $\bm{1}$   &  $\sfrac{2}{3}$  &      $1$      &   $\bm{3}$           &   Brane         \\
   $e^c$                &       $\bm{1}$     &    $\bm{1}$   &     $  2   $     &           $1$       &   $\bm{3}$   &   Brane           \\
   $Q$                  &        $\bm{3}$     &    $\bm{2}$  &  $\sfrac{1}{3}$  &       $1$        &   $\bm{3}$     &   Brane           \\
   $u^c_{1,2,3}$   &        $\bm{3}$     &    $\bm{1}$  & $ -\sfrac{4}{3}$ &       $-1$         &   $\bm{1''}, \bm{1'}, \bm{1}$     &    Bulk        \\
   $F$                  &         $\bm{1}$    &     $\bm{1}$  & $ 0       $          &            $i$       &  $\bm{3}$   &    Brane          \\ \hline
   $H_u$              &        $\bm{1}$     &    $\bm{2}$   &    $1    $           &         $-1$        &  $\bm{3}$  & Brane       \\
   $H_d$             &         $\bm{1}$     &    $\bm{2}$   &   $-1   $           &            $1$        &  $\bm{3}$   &      Brane         \\
   $\eta$              &         $\bm{1}$     &    $\bm{2}$  &   $ 1  $             &              $-i$       &  $\bm{1}$  &    Brane          \\ 
   $\sigma$         &         $\bm{1}$      &    $\bm{1}$  &  $   0    $          &              $-1$      &    $\bm{3}$   &      Bulk      \\  \hline 
\end{tabular}
\caption{Field representation content and symmetries of the model.}
\label{tab:FieldContent}
\end{table}

Given the symmetries defining this model, one can write the most general effective Yukawa Lagrangian after compactification, i.e.
at an energy regime much lower than the compactification scale. We write the Yukawa terms for the different scalars separately to simplify the analysis.
Notice that the bold subscripts of the terms below indicate its transformation properties under the remnant $A_4$ family symmetry.
The explicit expressions for the invariant multiplet products are written in Appendix \ref{app:a4}.

The Yukawa interaction terms of down-type quarks and charged leptons all have the same structure, given by
\begin{equation}
\mathcal{L}^{\scriptscriptstyle{\text{Yukawa}}}_{\scriptstyle{H_d}}=y^{e}_1 \left( LH_de^c\right)_{\bm{1}_1} + y^e_2  \left(LH_de^c\right)_{\bm{1}_2} + y^{d}_1 \left(QH_dd^c\right)_{\bm{1}_1} + y^d_2  \left(QH_dd^c\right)_{\bm{1}_2} + \text{H.c.},
\label{eq:HdYukawa}
\end{equation}
whereas the transformation properties of the up-type quark fields under $A_4$ yield the following Yukawa terms:
\begin{equation}
\mathcal{L}^{\scriptscriptstyle{\text{Yukawa}}}_{\scriptstyle{H_u}}=y^{u}_1 \left( QH_u\right)_{\bm{1'}}u^c_1 + y^{u}_2 \left( QH_u\right)_{\bm{1''}}u^c_2 +y^{u}_3 \left( QH_u\right)_{\bm{1}}u^c_3 + \text{H.c.}
\label{eq:HuYukawa}
\end{equation}
The dark fermion triplet $F$ couples to the scalar field $\sigma$. The latter acquires a vacuum expectation value which drives the spontaneous breaking of lepton number symmetry, 
$\mathbb{Z}_4$, and the $A_4$ family symmetry, giving rise to Majorana mass terms for the dark fermions. 
\begin{equation}
\mathcal{L}^{\scriptscriptstyle{\text{Yukawa}}}_{\scriptstyle{\sigma}} =y^{\sigma} \left( F^{T}F \sigma\right)_{\bm{1}_{1}} + \text{H.c.}
\label{eq:SigmaYukawa}
\end{equation}
The dark scalar $\eta$ plays a crucial role in the model since it couples with both dark fermions and neutrinos 
\begin{equation}
\mathcal{L}^{\scriptscriptstyle{\text{Yukawa}}}_{\scriptsize{\eta}}= y^{\eta}_1 \left( L  \eta F \right)_{\bm{1}} + \text{H.c.}
 \label{eq:EtaYukawa}
\end{equation}
In the following, we will assume that all Yukawa couplings are real, and therefore that the model preserves a trivial CP symmetry.  

The scalar potential $ V(H_d,H_u,\eta, \sigma)$ comprises all terms up to quartic interactions consistent with all the symmetries of the model.
It contains enough freedom in parameter space to drive the spontaneous breaking of the gauge symmetries down to $\mathrm{U(1)_{EM}}$. 
In this work, we will not perform a detailed analysis of the scalar potential of the model, since most of its properties are not relevant for the phenomenological studies of this paper.  
There is, however, a term of  special importance in the scalar potential for the neutrino mass generation mechanism, namely
\begin{equation}
V(H_{u},H_{d}, \eta) \supset \frac{1}{2} \lambda_{5} \left[ \left( {H_{d}}^{T} \left(i\sigma_{2}\right)\eta \right)_{\bm{3}} \left( H_{u}^{\dagger} \eta\right)_{\bm{3}} \right]_{\bm{1}} + \text{H.c.}, 
\label{eq:lambda5terms}
\end{equation}
with coupling constant $\lambda_5$ and $\sigma_{2}$ as the second Pauli matrix. 
This term is responsible for lifting the degeneracy of the mass eigenstates of the neutral components of $\eta$, which we shall denote $\sqrt{2}\text{Re}(\eta^0)$ and $\sqrt{2}\text{Im}(\eta^0)$.
The interplay between the dark and scalar sectors will generate the neutrino mass as a scotogenic correction at the one-loop level, as illustrated in Fig.~\ref{fig:ScotoLoop}.

The symmetry breakdown of the model is performed in two steps. To start with, at high energies the electroweak singlet scalar field $\sigma$ develops a vacuum expectation value (VEV) compatible with the extra-dimensional boundary conditions.
Subsequently, at lower energies, the Higgs iso-doublets $H_u, H_d$ acquire corresponding VEVs according to the minimization of the scalar potential.  Let us take a closer look at the high scale $A_4$ symmetry breaking produced by $\sigma$.

We introduce a boundary condition $P$, consistent with the orbifold construction. It defines a non-trivial gauge/Poincar\'e twist of the orbifold, and must be a symmetry transformation of the Lagrangian.
We assume that the transformation $P$ acts trivially on $A_4$ singlet bulk fields.
Therefore, the only field in the bulk that transforms non-trivially under $P$ is the flavour bulk triplet $\sigma$, which for consistency with Eq.~(\ref{eq:id3c}) must comply with the boundary condition.
\begin{equation}
\sigma(x,z)=P \sigma(x,-z).
\label{eq:bcOfSigma}
\end{equation}
From the six-dimensional Lagrangian, the invariance of the kinetic term of the $\sigma$ field requires $P \in SU(3)$, while the condition in Eq. (\ref{eq:bcOfSigma}) ensures the matrix $P$ will leave invariant the interactions of fields in the brane.
Therefore the boundary condition matrix must satisfy
\begin{equation}
P \in SU(3), \qquad P^2= \bm{1}_{\scriptsize 3\times 3}, \quad P^{\dagger} = P.
\label{eq:Pcon}
\end{equation}
Note that, while in general the matrix $P \in \mathrm{SU(3)}$, the full Lagrangian only has an $A_4$ symmetry. Therefore the matrix $P$ can't be $\mathrm{SU(3)}$ rotated into a trivial form.

The boundary condition of the $\sigma$ field applies also to its VEV alignment. Therefore in this model, the masses of dark fermions $F$ are a direct outcome of the boundary condition of the bulk field $ \sigma $ in the two extra dimensions
\begin{equation}
\braket{\sigma}=P \braket{\sigma}.
\label{eq:bcOfSigmaVEV}
\end{equation}
The boundary condition matrix $P$ is a property of the orbifold and therefore it is arbitrary. In the present work we will not assume a particular form for $P$. Instead, we will adopt the most general VEV alignment compatible with the spontaneous breaking of lepton number and the $A_4$ family symmetry, that (up to unphysical rephasings) can be expressed as
\begin{equation}
\braket{\sigma}= v_{\sigma}\left( \begin{array}{c} \epsilon_1^{\sigma} \hspace{2pt} e^{i\varphi} \\ \epsilon_2^{\sigma} \\ 1 \end{array} \right), \qquad \text{with} \qquad  v_{\sigma}, \epsilon_1^{\sigma},  \epsilon_2^{\sigma} \in \mathbb{R} \quad \text{and} \quad 0\leq \varphi < \pi.
\label{eq:SigmaVEVBoundary}
\end{equation}
The parametrization of the respective boundary condition matrix $P$ in terms of ($\epsilon^\sigma_{1,2},\varphi$) is written explicitly in Appendix  \ref{app:Pmatrix}.

Since the $A_4$ symmetry is broken spontaneously  at the $v_{\sigma}$ scale, in the second stage of spontaneous symmetry breaking, we assume that the weak iso-doublets $H_u$ and $H_d$ obtain the most general $A_4$ breaking VEVs,
compatible with the preservation of trivial CP, and parameterized as 
\begin{equation}\label{vevs}\begin{split}
\braket{H_u}=v_u\left(\begin{array}{c}\epsilon_1^u\\ \epsilon_2^u \\ 1\end{array}\right),\ \ \ \braket{H_d}=v_d \left(\begin{array}{c}\epsilon_1^d \\ \epsilon_2^d \\ 1\end{array}\right),
\end{split}\end{equation}
with real parameters $v_u,v_d,\epsilon^{u,d}_{1,2}$.

\subsection{Phenomenology from Extra Dimensions}

The assumption of Extra Dimensions implies the existence of a collection of infinitely many 4-D fields associated with every field in the bulk, called a Kaluza-Klein (KK) tower, with masses $(n^2+m^2)M_c$ determined by positive integers $n,m$.
In our model, the fields in the bulk are the \SM gauge fields $g_\mu, W_\mu, B_\mu$, the right-handed quarks $u_i^c$ and the SM singlet scalar $\sigma$.
These massive fields generate a rich phenomenology that can be tested in current colliders.
Notice that all the phenomenological analysis performed in this paper is independent of the compactification scale $M_c$. 

One of the most well-known predictions derived from the existence of a KK tower is the generation of flavour Changing Neutral Currents (FCNCs).
These arise when the extra-dimensional fields in the bulk are allowed to have an explicit mass term, preventing the simultaneous flavour diagonalization of their higher KK modes with the zeroth level Lagrangian. \cite{Cheung:2001mq,Barbieri:2004qk}.
Since our model is built in six flat dimensions, the bulk fermions are 6-D chiral fields, with no explicit mass term and therefore free of FCNCs by construction.\\[-.4cm] 

We now turn to Electroweak Precision observables. In our model, the most important effect comes from the contribution of the KK modes to the Peskin-Takeuchi parameters $S$, $T$ and $U$,
which are modified mainly by the tower of massive vector $\mathrm{SU(2)_L}$ triplets.
In contrast, the massive vector KK singlet $B_\mu$, as well as the right-handed quarks, have much less impact on these observables.
The current experimental bound for our setup (2 Extra non-Universal Dimensions) is $M_c>2.1\ {\rm TeV}$ \cite{Deutschmann:2017bth,Ganguly:2018pzs}.
If the compactification scale is sufficiently close to $2\ {\rm TeV}$, the electroweak precision tests could in principle probe the extra dimensions in the High Luminosity LHC run. 

\section{Fermion Masses}
\label{sec:fermion-masses}

After spontaneous symmetry breaking, the mass matrices of quarks and charged leptons become 
\begin{equation}\begin{split}
 M_u&=v_u\left(\begin{array}{ccc} y_1^u\epsilon_1^u &y_2^u \epsilon_1^u & y_3^u\epsilon_1^u \\
 y_1^u\epsilon_2^u \omega^2&  y_2^u\epsilon_2^u \omega &   y_3^u\epsilon^u_2 \\
  y_1^u \omega& y_2^u \omega^2&y_3^u\end{array}\right),\\
M_d&=v_d\left(\begin{array}{ccc} 0 & y_1^d\epsilon_1^d & y_2^d \epsilon_2^d \\
 y_2^d\epsilon_1^d & 0 &  y_1^d\\
 y_1^d\epsilon_2^d & y_2^d&0\end{array}\right),\\
M_e&=v_d\left(\begin{array}{ccc} 0 & y_1^e\epsilon_1^d & y_2^e \epsilon_2^d \\
 y_2^e\epsilon_1^d & 0 &  y_1^e\\
 y_1^e\epsilon_2^d & y_2^e&0\end{array}\right),
 \label{eq:massmat}
\end{split}\end{equation}
 where all Yukawa couplings are assumed to be real due to our imposition of trivial CP symmetry.

\begin{figure}[H]
\centering
\includegraphics[width=0.5\textwidth]{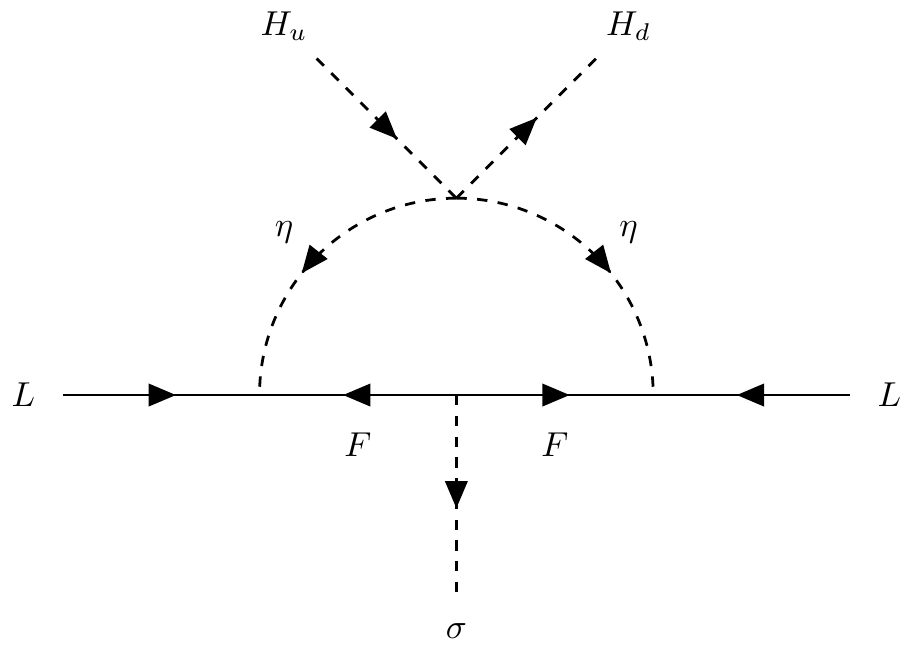}
\caption{One-loop diagram for Majorana neutrino masses, mediated by the ``dark sector'' particles.}
\label{fig:ScotoLoop}
\end{figure}

Likewise, the $A_4$ flavour symmetry structure of $\mathcal{L}^{\scriptscriptstyle{\text{Yukawa}}}_{\scriptstyle{\sigma}}$ in Eq. (\ref{eq:SigmaYukawa}) implies that the Majorana mass matrix of the dark fermions $M_{F}$ must have the following structure
\begin{equation}
M_F=y_{\sigma} v_{\sigma}\left(\begin{array}{ccc} 0 &1 &  \epsilon_2^{\sigma}  \\
 1 & 0 &  \epsilon_1^{\sigma} e^{i\varphi} \\
  \epsilon_2^{\sigma} &   \epsilon_1^{\sigma} e^{i\varphi}& 0 \end{array}\right).
\label{eq:DarkFermionsMassMatrix}
\end{equation}
In order to describe our one-loop scotogenic mechanism for neutrino masses we write the dark fermion $F$ fields in the mass eigenstate basis ($\tilde{F}$). 
This can be done by performing the singular value decomposition of the dark fermion mass matrix $M_F$. Since the latter is symmetric, only one unitary matrix $V$ is needed in the Takagi decomposition~\cite{Schechter:1980gr},
\begin{equation}
y^{\sigma} \left( F^{T}F \sigma\right)_{\bm{1}_{1}}= F^{T} M_{F} F=F^{T} V^{T} D V F=\left(VF \right)^{T} D \left( VF\right) \equiv {\tilde{F} }^{T} D \tilde{F},
\label{eq:ChangeBasisDarkFermions}
\end{equation}
where $\tilde{F} \equiv VF$ is the dark fermion triplet written in the mass eigenstate basis and $D=\mathrm{diag}(m_{F_1},m_{F_2},m_{F_3})$. We can then rewrite Eq. (\ref{eq:EtaYukawa}) as
\begin{equation}
\mathcal{L}^{\scriptscriptstyle{\text{Yukawa}}}_{\scriptsize{\eta}}= y^{\eta}_1\eta  \left( L  V^{\dagger}\tilde{F} \right) + \text{H.c.}
 \label{eq:EtaYuakwaMass}
\end{equation} 
As already noted, in this model neutrino masses are forbidden at tree-level due to the auxiliary $\mathbb{Z}_4$ symmetry. 
However, thanks to the mediation of the dark fields $\eta$ and $F$, neutrino masses emerge at one-loop through the diagram depicted in Fig. \ref{fig:ScotoLoop}.  
The resulting neutrino mass matrix has the basic scotogenic structure~\cite{Ma:2006km}.

Defining $y^{\eta}_1V^{\dagger} \equiv h$ in Eq. (\ref{eq:EtaYuakwaMass}) we can write the expression for the one-loop neutrino mass matrix  $M_{\nu}$ as
\begin{equation}
\left(M_{\nu}\right)_{ij} = \sum^3_{k} \frac{h_{ik} (h^{T})_{kj}}{16\pi^{2}} S(m_{F_k}),
 \label{eq:NuMassMatrix}
\end{equation} 
where $S(m_{F_k})$ stands for the loop factor
\begin{equation}
 S(m_{F_k})= m_{F_k} \left( \frac{m^2_R }{m^2_R- m^2_{F_k}}  \ln \frac{m^2_R}{m^2_{F_k}}- \frac{m^2_I }{m^2_I- m^2_{F_k}}  \ln \frac{m^2_I}{m^2_{F_k}}\right),
 \label{eq:ScotoFactor}
\end{equation} 
with $m_{R}=m(\sqrt{2}\hspace{2pt} \text{Re}  \hspace{2pt} \eta^{0})$, $m_{I}=m(\sqrt{2} \hspace{2pt} \text{Im}   \hspace{2pt} \eta^{0})$ and
\begin{equation}
m^{2}_{R} - m^{2}_{I} \equiv  2 \lambda_5  \left( \braket{H_u}_{\bm{3}} \braket{H_d}_{\bm{3}} \right)_{\bm{1}}.
\label{eq:lambda5Prop}
\end{equation}
Neutrino masses are not only loop-suppressed but also symmetry-protected, as it vanishes when $\lambda_5 \to 0$, see Eq.~(\ref{eq:lambda5terms}). 

After spontaneous symmetry breaking, the auxiliary $\mathbb{Z}_4$ breaks down to a preserved $\mathbb{Z}_2$ that stabilizes the lightest dark particle, as in \cite{Ma:2006km}.
The model has two possible dark matter candidates: the lightest neutral scalar contained in $\eta$, or the lightest Majorana fermion in $F$.
In both cases, the phenomenology of dark matter generically coincides with that of the scotogenic scenario.
For further details on the current status of these dark matter candidates see \cite{Avila:2021mwg} and references therein. \\[-.3cm]

Apart from its scotogenic nature, there are additional neutrino mass features that arise from our family symmetry. 
In the next section, we discuss the flavour predictions of our model, both in the quark and lepton sectors.  
In particular,  we stress that our proposal will be tested in many ways, especially by the measurement of neutrino oscillation parameters and the improved determination of
down-type quark and charged lepton masses.

\section{ flavour predictions}
\label{sec:analysis}

Our model successfully reproduces the observed pattern of fermion masses and mixings from the $A_4$  family symmetry that results from the orbifold compactification of the extra dimensions.
The next subsections describe the main flavour predictions of our model as well as some of the resulting phenomenological implications, i.e.
\begin{itemize}
\item The ``golden'' mass relation amongst down-type quark and charged lepton masses,
\item a successful global description of all flavour observables,
\item the mass ordering, atmospheric octant, and \znbb prediction, as well as
\item predictions for the neutrino oscillation parameters and leptonic CP violation.
\end{itemize}

\subsection{Golden Relation}
\label{sec:golden}

A central prediction of our model is a direct consequence of the charge assignments of the down-type quarks and the charged leptons under the $A_4$ flavour symmetry. 
Indeed, after spontaneous symmetry breaking these fields get masses from the same Higgs doublet $H_d$, leading to the mass matrices in Eq.(\ref{eq:massmat}).
After diagonalization to physical states, one obtains the golden quark-mass relation
\begin{equation}\label{golden}
\frac{m_\tau}{\sqrt{m_\mu m_e}}\approx\frac{m_b}{\sqrt{m_s m_d}},
\end{equation}
This relation emerges in a broad class of flavour models~\cite{Morisi:2011pt,King:2013hj,Morisi:2013eca,Bonilla:2014xla,Reig:2018ocz}.
However, in the present case, the underlying $A_4$ family symmetry is dictated by the orbifold compactification of the original 6-dimensional theory~\cite{deAnda:2019jxw}.
Given the current experimental measurements of the relevant masses, the golden relation holds with good precision. 
Besides, the golden formula is a very robust prediction under the renormalization group evolution, since it involves only ratios of fermion masses.

\subsection{Global Flavour Fit}
\label{sec:global-flavour-fit}

In the following analysis, we adopt the symmetrical parametrization of the fermion mixing matrices~\cite{Rodejohann:2011vc}.
In the quark sector, choosing the PDG ordering prescription leads to the standard form for the Cabibbo-Kobayashi-Maskawa (CKM) matrix characterizing quark mixing~\cite{Zyla:2020zbs}
\begin{equation}  \label{eq:CKM}
 V_{CKM}= \left( 
\begin{array}{ccc}
c^q_{12} c^q_{13} & s^q_{12} c^q_{13}  & s^q_{13} e^{-i\delta^q}
\\ 
-s^q_{12} c^q_{23}- c^q_{12} s^q_{13} s^q_{23} e^{ i \delta^q} & c^q_{12} c^q_{23} - s^q_{12} s^q_{13} s^q_{23} e^{ i \delta^q } & c^q_{13} s^q_{23} \\ 
s^q_{12} s^q_{23} - c^q_{12} s^q_{13} c^q_{23} e^{ i\delta^q } & - c^q_{12} s^q_{23}  - s^q_{12} s^q_{13} c^q_{23} e^{i\delta^q } & c^q_{13} c^q_{23}%
\end{array}
\right)\,, 
\end{equation}
On the other hand, for the case of leptons we get
\begin{equation}  \label{eq:symmetric_para}
 K= \left( 
\begin{array}{ccc}
c^{\ell}_{12} c^{\ell}_{13} & s^{\ell}_{12} c^{\ell}_{13} e^{ - i \phi_{12} } & s^{\ell}_{13} e^{ -i \phi_{13} }
\\ 
-s^{\ell}_{12} c^{\ell}_{23} e^{ i \phi_{12} } - c^{\ell}_{12} s^{\ell}_{13} s^{\ell}_{23} e^{ -i ( \phi_{23} -
\phi_{13} ) } & c^{\ell}_{12} c^{\ell}_{23} - s^{\ell}_{12} s^{\ell}_{13} s^{\ell}_{23} e^{ -i ( \phi_{23} +
\phi_{12} - \phi_{13} ) } & c^{\ell}_{13} s^{\ell}_{23} e^{- i \phi_{23} } \\ 
s^{\ell}_{12} s^{\ell}_{23} e^{ i ( \phi_{23} + \phi_{12} ) } - c^{\ell}_{12} s^{\ell}_{13} c^{\ell}_{23} e^{ i
\phi_{13} } & - c^{\ell}_{12} s^{\ell}_{23} e^{ i \phi_{23} } - s^{\ell}_{12} s^{\ell}_{13} c^{\ell}_{23} e^{
-i ( \phi_{12} - \phi_{13} ) } & c^{\ell}_{13} c^{\ell}_{23}%
\end{array}
\right)\,,
\end{equation}
where $c^{f}_{ij}\equiv \cos\theta^f_{ij}$ and $s^{f}_{ij}\equiv \sin\theta^f_{ij}$. This description is exactly the same as the early proposal in~\cite{Schechter:1980gr}, supplemented by the convenient PDG factor ordering convention.
However, this symmetrical description of lepton mixing provides a neater description of leptonic CP violation than the PDG form. 
Indeed, in the symmetrical parameterization the Dirac CP violating phase that enters in neutrino oscillations is identified with the ``rephasing-invariant'' combination
\begin{equation}
\delta^{\ell}=\phi_{13}-\phi_{12}-\phi_{23}.
\end{equation}
This is the leptonic analogue of the quark Jarlskog parameter.  
On the other hand, the effective mass parameter characterizing the amplitude for neutrinoless double beta decay involves only the two Majorana phases~\cite{Rodejohann:2011vc},
\begin{equation}\label{eq:n0bb}
\langle m_{\beta\beta}\rangle=\left|\sum_{j=1}^3 K_{ej}^2 m_j\right|=
\left|c^{\ell\,2}_{12}c^{\ell\,2}_{13} m_1 + s^{\ell\,2}
_{12}c^{\ell\,2}_{13} m_2 e^{2i\phi_{12} }+ s^{\ell\,2}_{13} m_3 e^{2i\phi_{13}}\right|.
\end{equation}
One sees how this symmetrical presentation provides a more transparent description of leptonic CP violation and its impact on the \znbb amplitude.  \\[-.2cm]

We will now show that, due to the reduced number of parameters available, our model makes other interesting flavour predictions.
Indeed, we have 16 independent parameters characterizing the flavour sector, identified as follows: 
8 real Yukawa couplings $y^{e,d}_{1,2}$, $y^u_{1,2,3}$, $y^{\eta}_1$, 6 real VEV ratios $\epsilon^{u,d}_{1,2}$, $\epsilon^\sigma_{1,2}$, one quartic coupling $\lambda_5$ and one CP violating phase $\varphi$ contained in $\vev{ \sigma}$.
These parameters describe 22 observables, namely, 12 masses, 4 CKM parameters, plus 6 lepton mixing matrix parameters including the 2 Majorana phases.
As a result,  there are in total, 6 flavour predictions in the model at low energies, one of which we readily identify with the golden quark-lepton mass relation in Eq.(\ref{golden}). \\[-.4cm]

An interesting feature of our model is that $\varphi$ is the only source of CP-violation, which comes directly from the orbifold boundary condition. 
The single phase $\varphi$ must be fixed to reproduce the well measured value of the quark sector CP violating phase $\delta^q$.
Once fixed the value of $\varphi$, there is no freedom in choosing the three CP violating phases of the lepton sector $\phi_{12,13,23}$. 
Thus, there are three flavour predictions in the model related to CP violation in the lepton sector. \\[-.4cm]

The fifth prediction of the model can be identified with the mass of the lightest neutrino, as we show below.
Finally, the sixth is a correlation between the mixing angles of the lepton mixing matrix. This last prediction is not evident from the global flavour fit discussed below. 
In an effort to identify the nature of this prediction we will perform a further exploration of the parameter space around the global minimum of the fit. \\[-.4cm]

In order to extract the predictions of the model, we perform a global flavour fit that fixes the values of the model parameters using the available experimental flavour data.
The fit has been performed by scanning the values of the 16 real independent model parameters that yield 22 flavour observables (6 lepton masses, 3 lepton mixing angles, 3 lepton CP phases, 6 quark masses,
3 CKM mixing angles and the CKM phase). 
The global flavour fit to the available experimental data is performed by minimizing the chi-square function defined as
\begin{equation}
\chi^2=\sum (\mu_{\text{exp}}-\mu_{\text{model}})^2/\sigma^{2}_{\text{exp}},
\end{equation}
where the sum runs through the 19 measured physical parameters, i.e. 6 quark masses, 3 CKM mixing angles, 1 CKM CP phase,
3 charged lepton masses, 3 lepton mixing angles, 1 lepton CP violating phase and 2 neutrino squared mass splittings. 
Note that we have only limits on the lightest neutrino mass from experiment and, at the moment, no information on the Majorana \znbb phases.

In our numerical minimization of the chi-square function with respect to the 16 independent model parameters, all quark and charged-lepton masses were evaluated at the same energy scale,
which we choose to be $M_Z$ \cite{Antusch:2013jca}. This assumption has been shown to be compatible with the golden relation in Ref.~\cite{deAnda:2019jxw}. 
For our study, the neutrino oscillation parameters have been extracted from the global fit \cite{deSalas:2020pgw,zenodo}, a choice justified by the fact that the effect on neutrino (and quark)
parameters induced by the running to $M_Z$ is negligible \cite{Xing:2007fb,Antusch:2013jca}. In the same vein, the rest of the observables have been taken from the PDG \cite{Zyla:2020zbs}.
We have made use of the Mathematica Mixing Parameter Tools package~\cite{Antusch:2005gp} in the extraction of the flavour physical observables from the mass matrices in Eq.(\ref{eq:massmat}).

The results of our flavour fit are summarized in Table \ref{tab:fit}. 
The minimum sits at $\chi^2\approx 2$, showing that the model successfully reproduces the observed pattern of fermion masses and mixing.
Besides, from the results in Table \ref{tab:fit} one can read directly the central predictions of the model regarding the mass of the lightest neutrino and the CP violating phases of the lepton sector. 
In order to study in more detail the predictions of the theory we have performed a further exploration of the parameter space by randomly varying the model parameters around the global minimum in Table \ref{tab:fit} 
within a range that covers the experimentally allowed $3\sigma$ ranges of all measured flavour observables. 
In the following subsections, we will discuss the resulting predictions associated with this parameter region. 

\begin{table}[ht]
	\centering
	\footnotesize
	\renewcommand{\arraystretch}{1.1}
	\begin{tabular}[t]{|lc|r|}
		\hline
		Parameter &\qquad& Value \\ 
		\hline
		$y^e_1v_d/\mathrm{GeV}$ &\quad& $-1.745$ \\
		$y^e_2v_d/(10^{-1}\mathrm{GeV})$ && $1.021$ \\
		\rule{0pt}{3ex}%
		$y^d_1v_d/(10^{-2}\mathrm{GeV})$ &&$-5.039$ \\
		$y^d_2v_d /\mathrm{GeV}$ && $2.852$ \\
		\rule{0pt}{3ex}%
		\rule{0pt}{3ex}%
		$y^u_{1} v_u/(10^{-1}\mathrm{GeV})$ &&$6.074$ \\
		$y^u_2v_u/(10^2\mathrm{GeV})$ && $1.712$ \\
		$y^u_3v_u/\mathrm{GeV}$ && $7.157$ \\
		\rule{0pt}{3ex}%
		$\epsilon^u_1/10^{-4}$ && $7.055 $ \\
		$\epsilon^u_2/10^{-2}$ && $-5.044$ \\
		\rule{0pt}{3ex}%
		$\epsilon^d_1/10^{-3}$ && $-2.814$ \\
		$\epsilon^d_2/10^{-3}$ && $5.833$ \\
		\rule{0pt}{3ex}%
		$\epsilon^{\sigma}_1$ && $1.501$\\
		$\epsilon^{\sigma}_2$ && $-0.654$\\
		$\varphi$ && $3.527$\\
		$(y^{\eta}_{1})^2 y_{\sigma} v_{\sigma}/(\mathrm{KeV})$ && $1.813$\\
		$2 \lambda_5  \braket{H_u}  \braket{H_d} /(\mathrm{KeV})^2$ && $0.012$\\
		\hline	
	\end{tabular}  
	\hspace*{0.5cm}
	\begin{tabular}[t]{ |l |c|c c |c| }
		\hline
		\multirow{2}{*}{Observable}& \multicolumn{2}{c}{Data} & & \multirow{2}{*}{Model best fit}  \\
		\cline{2-4}
		& Central value & 1$\sigma$ range  &   & \\
		\hline
		$\theta_{12}^\ell$ $/^\circ$ & 34.3 & 33.3 $\to$ 35.3 && $33.0$  \\ 
		$\theta_{13}^\ell$ $/^\circ$ & 8.45 & 8.31 $\to$ 8.61  && $8.52$  \\  
		$\theta_{23}^\ell$ $/^\circ$ & 49.26 & 48.47 $\to$ 50.05  && $50.44$ \\ 
		$\delta^\ell$ $/^\circ$ & 194 & 172 $\to$ 218 && $192$  \\
		$m_e$ $/ \mathrm{MeV}$ & 0.486 &  0.486 $\to$ 0.486 && $0.486$ \\ 
		$m_\mu$ $/  \mathrm{GeV}$ & 0.102 & 0.102  $\to$ 0.102  &&  $0.102$ \\ 
		$m_\tau$ $/ \mathrm{GeV}$ &1.745 & 1.743 $\to$1.747 && $1.745$ \\ 
		$\Delta m_{21}^2 / (10^{-5} \, \mathrm{eV}^2 ) $ & 7.50  & 7.30 $\to$ 7.72 && $7.50$  \\
		$\Delta m_{31}^2 / (10^{-3} \, \mathrm{eV}^2) $ & 2.55  & 2.52 $\to$ 2.57 &&  $2.54$ \\
		$m_1$ $/\mathrm{meV}$  & & & & $135.35 $ \\ 
		$m_2$ $/\mathrm{meV}$  & && & $ 135.63$ \\ 
		$m_3$ $/\mathrm{meV}$  & && & $144.43 $ \\
		$ \phi_{12} $ $/^\circ$ & & && $87.01$  \\
		$ \phi_{13}$  $/^\circ$& & && $190.30$  \\    
		$ \phi_{23}$  $/^\circ$& & && $271.05$  \\    	
		\hline
		$\theta_{12}^q$ $/^\circ$ &13.04 & 12.99 $\to$ 13.09 &&  $13.04$ \\	
		$\theta_{13}^q$ $/^\circ$ &0.20 & 0.19 $\to$ 0.22 && $0.20$  \\
		$\theta_{23}^q$ $/^\circ$ &2.38& 2.32 $\to$ 2.44 && $2.38$  \\	
		$\delta^q$ $/^\circ$ & 68.75 & 64.25 $\to$ 73.25  & & $60.23$\\
		$m_u$ $/ \mathrm{MeV}$ & 1.28 & 0.76$\to$ 1.55 && $1.28$  \\	
		$m_c$ $/ \mathrm{GeV}$ & 0.626 & 0.607 $\to$ 0.645 &&  $0.626$ \\	
		$m_t$ $/\mathrm{GeV}$  	  & 171.6& 170 $\to$ 173  && $171.6$ \\
		$m_d$ $/ \mathrm{MeV}$ & 2.74 & 2.57 $\to$ 3.15 &&  $2.49$ \\	
		$m_s$ $/ \mathrm{MeV}$ & 54 & 51 $\to$ 57 && $54$ \\
		$m_b$ $/ \mathrm{GeV}$	  & 2.85 &  2.83 $\to$   2.88 &&  $2.85$\\
		\hline
		$\chi^2$ & & & & $1.96$ \\
		\hline		
	\end{tabular}
        \caption{Flavour parameters and observables: measured versus predicted values for the best fit point.} 
	\label{tab:fit}
\end{table}

\subsection{Neutrino Mass Ordering and Neutrinoless Double Beta Decay}
\label{sec:neutr-double-beta}

In our exploration of the parameter space of the model consistent at $3\sigma$ with all current experimental data we have found only Normal Ordering (NO) for the neutrino mass spectrum.
In particular, the best fit point in Table \ref{tab:fit}, is a solution for positive $\Delta m_{31}^2$, corresponding to NO, with a rather high absolute scale for the neutrino masses.

\begin{figure}[]
\centering
\includegraphics[width=0.5\textwidth]{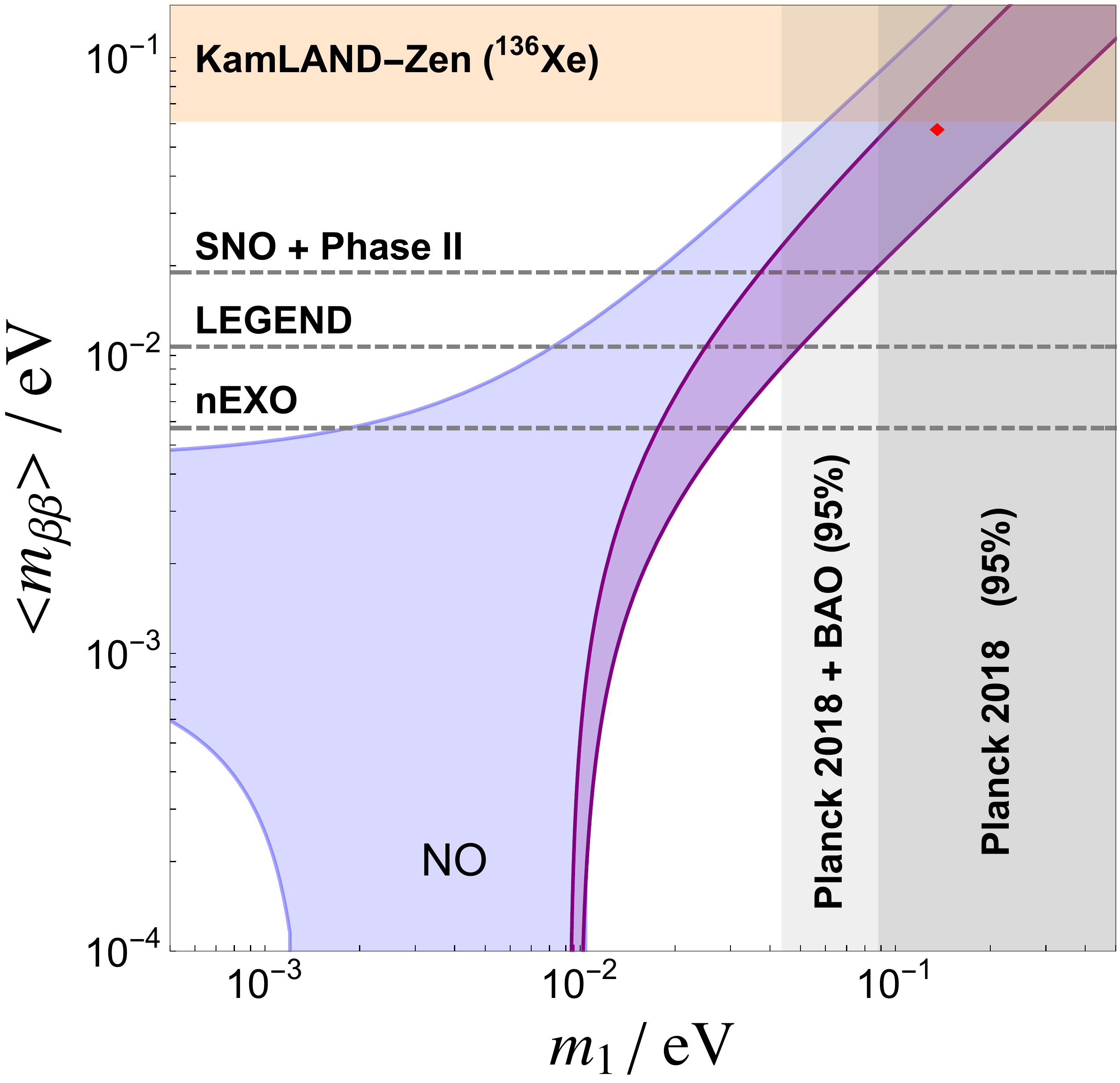}
\caption{
  Effective \znbb decay amplitude as a function of the lightest neutrino mass $m_1$. From the global analysis, we found that only normal ordering is allowed in our model.
  The blue region is the generic one consistent with oscillations at $2\sigma$. The purple region is the one allowed at $3\sigma$ around the global best fit point in Table \ref{tab:fit}, marked in red.
  The current KamLAND-Zen limit is shown in orange, and the projected sensitivities of upcoming experiments are indicated in dashed horizontal lines.
  The vertical gray bands represent the current sensitivity of cosmological data from the Planck collaboration (dark shade), and in combination with BAO data (light shade). }
\label{fig:neutrinoless1}
\end{figure}

Regarding neutrinoless double beta decay, plugging the CP predictions for the Majorana phases and the neutrino masses of Table \ref{tab:fit} into Eq.(\ref{eq:n0bb}), we obtain the following prediction for the effective amplitude that characterizes the process:
\begin{equation}
\vev{ m_{\beta\beta}}= 58.08 \,\mathrm{meV}.
\end{equation}
A more detailed analysis is presented in Figure \ref{fig:neutrinoless1}, where we plot in purple the region of predicted values for $\vev{ m_{\beta\beta}}$ as a function of the lightest neutrino mass $m_1$.  The value for the lightest neutrino mass at the best-fit point from our $\chi^2$ analysis is found to be $m_1=135.35 \text{ meV}$ as indicated in Table  \ref{tab:fit}. 
To be conservative, we randomly varied the model parameters within a range that covers the experimentally allowed $3\sigma$ range and the best fit point of Table \ref{tab:fit}, marked in red.
One sees that the predicted central value of $\vev{ m_{\beta\beta}}$ lies very close to the current experimental bound of Kamland-Zen $(61 - 165\; \mathrm{meV})$ \cite{KamLAND-Zen:2016pfg} shown as the top orange horizontal band in Fig.~\ref{fig:neutrinoless1}.
For comparison, we have also displayed the projected sensitivities of the next generation of \znbb experiments LEGEND \cite{Abgrall:2017syy}, SNO + Phase II \cite{Andringa:2015tza}, and nEXO \cite{Albert:2017hjq} as the dotted horizontal lines.

Notice that the central value of the lightest neutrino mass $m_1$ obtained from the global flavour fit is currently disfavored by the latest results of the Planck collaboration on the sum of light neutrino masses~\cite{Planck:2018vyg}.
Tension with cosmology is further enhanced if the cosmological bound is refined by the addition of Baryon Acoustic Oscillations (BAO) data, as studied in \cite{Lattanzi:2020iik} and depicted in shades of gray in Figure \ref{fig:neutrinoless1}.
However, as shown in the figure, beyond the central prediction of the model, there is a relatively wide region of parameters compatible at $3\sigma$ with all the measured flavour observables and consistent with the cosmological bounds. One concludes that the best fit point of the global flavour fit will be probed not only by cosmological observations \cite{Planck:2018vyg} but possibly also by future beta decay endpoint studies.

\subsection {Predictions For The Leptonic Dirac CP Phase and Mixing Angles}  
\label{sec:leptonic-cp-phase}

\begin{figure}[]
\centering
\includegraphics[width=0.45\textwidth]{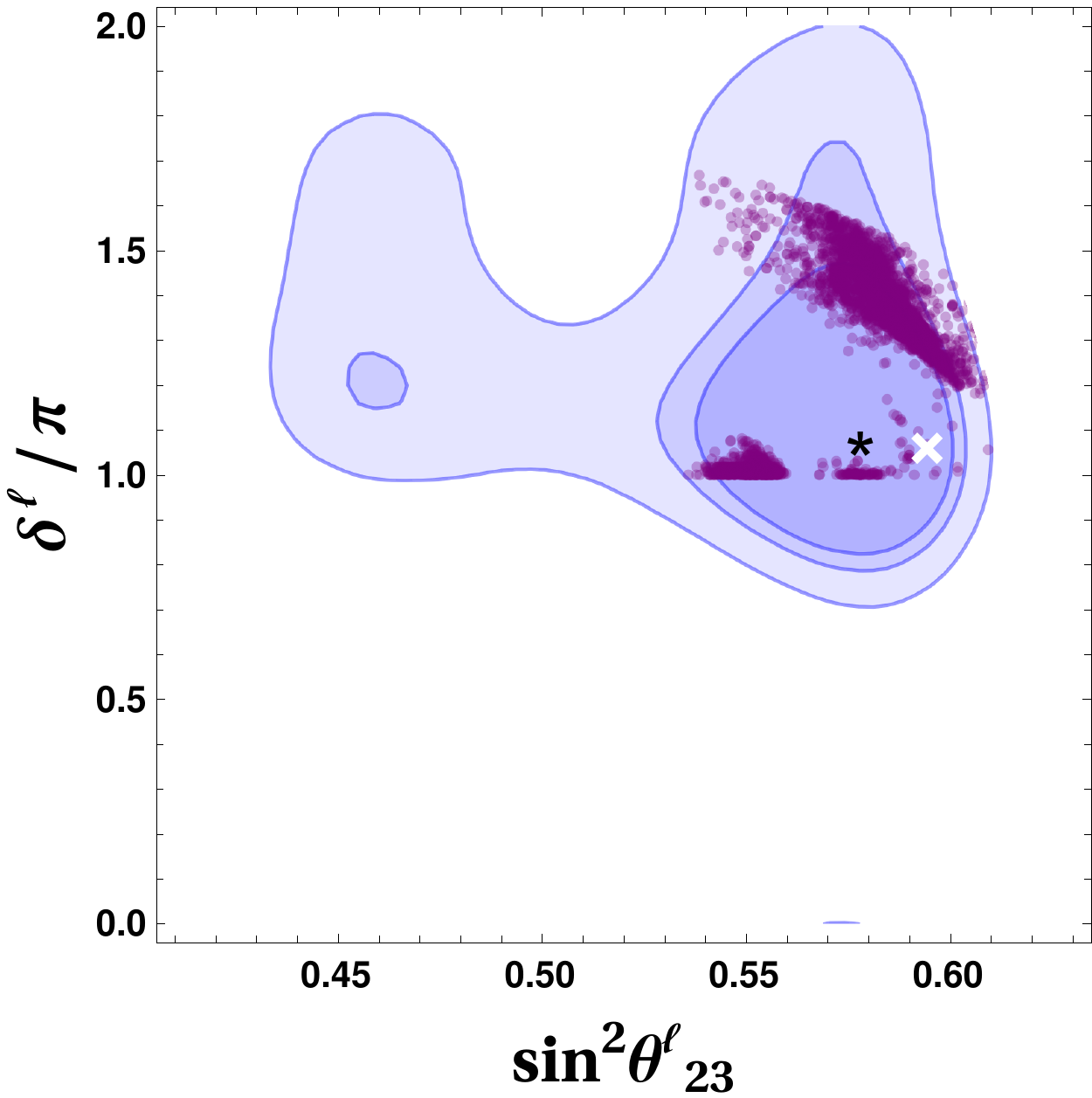}
\includegraphics[width=0.48\textwidth]{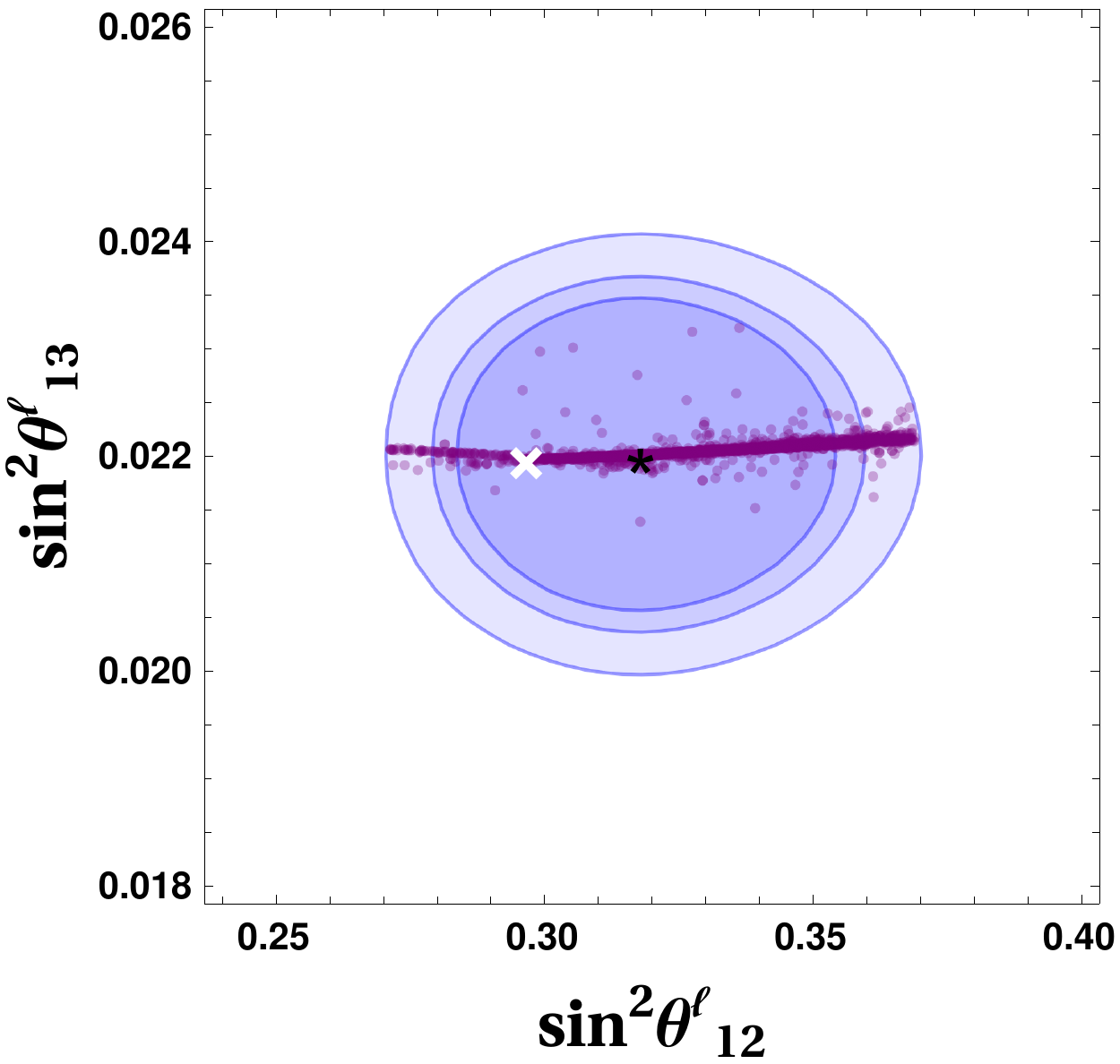}
\caption{
  Allowed values for the mixing angles and the leptonic Dirac CP phase.
  The purple points are compatible at $3\sigma$ with all the flavour observables, and the blue shades represent the 90, 95 and 99\% C.L. profiles of the Global oscillation fit \cite{deSalas:2020pgw,zenodo}.
  The black star stands for the central point of the global oscillation fit, and the white cross for the best fit point in Table \ref{tab:fit}.}
\label{fig:Lepton-angles-phase}
\end{figure}

In order to identify the predictions of the model concerning the oscillation angles and the Dirac CP phase of the lepton sector, we have explored the parameter space of the model by randomly varying the parameters
around the global best fit point in Table \ref{tab:fit}, while requiring compatibility with all measured flavour observables at the $3\sigma$ level. 
The results of this analysis are presented in Fig \ref{fig:Lepton-angles-phase} where the blue contours represent the 90, 95, and 99\% C.L. profiles from the global oscillation fit in \cite{deSalas:2020pgw, zenodo},
while the purple dots are compatible at $3\sigma$ with all experimental data. 
The best fit point of the global oscillation fit is marked with a black star, while that of the global flavour fit is indicated by a white cross. 
From our search of model parameters leading to values for all the listed observables inside their $3\sigma$ region, we find the following values for $\delta^l$ and $\sin^2\theta^l_{13}$
  at the extrema
\begin{align} 
\delta^l_{\text{min}} / \pi &= 1.0 \hspace{1pt} \text{,} \qquad \delta^l_{\text{max}} / \pi = 1.67,
\\ 
\sin^2\theta^l_{13\text{, min}} /10^{-2} &= 2.14 \hspace{1pt} \text{,} \qquad  \sin^2\theta^l_{13\text{, max}}/ 10^{-2}= 2.40.
\end{align} 
In Fig. \ref{fig:Lepton-angles-phase} one sees that the model predicts values of the leptonic Dirac CP phase restricted to the range $\delta^\ell \geq \pi$
and values of the atmospheric angle $\theta^\ell_{23}$ in the higher octant. 
Besides, the model makes a sharp prediction for the value of the reactor angle $\theta^\ell_{13}$ close to the central value in Table \ref{tab:fit}.  

\section{Summary and conclusions}
\label{sec:Conclusions}

Building upon previous work~\cite{deAnda:2019jxw,deAnda:2020pti} we have advocated the use of extra space-time dimensions as a promising approach to theories of flavour.
We have adopted a 6-dimensional orbifold construction in which the 4-dimensional family symmetry is dictated by the compactification to be the $A_4$ group, which we used to stack the three observed particle families,
except for the right-handed up-type quarks,  Table~\ref{tab:FieldContent}. 

In our bottom-up approach we focused on the low-energy phenomenology and using the 6-dimensional setup to motivate the 4-dimentional family symmetry structure.
The model naturally predicts the ``golden'' quark-lepton mass relation, Eq.~(\ref{golden}), providing a very good global description of all flavour observables, Table~\ref{tab:fit},
making also predictions for neutrino mass ordering, atmospheric octant, and neutrinoless double beta decay, Fig.~\ref{fig:neutrinoless1}.
Moreover, we made predictions for the neutrino oscillation parameters and leptonic CP violation, Fig.~\ref{fig:Lepton-angles-phase}.
Another important feature of our construction is that, on top of a predictive orbifold theory of flavour, we have implemented a scotogenic Majorana neutrino mass mechanism,
where WIMP dark matter candidates can be identified as the mediators of neutrino mass generation, and stabilized due to the presence of an auxiliary $\mathbb{Z}_4$ symmetry.

\acknowledgements 
\noindent

Work supported by the Spanish grants PID2020-113775GB-I00 (AEI / 10.13039/501100011033), Programa Santiago Grisolía (GRISOLIA/2020/025), and PROMETEO/2018/165 (Generalitat Valenciana) and
Fundacão para a Ciência e a Tecnologia (FCT, Portugal) through grant CERN/FIS-PAR/0004/2019.
CAV-A is supported by the Mexican C\'atedras CONACYT project 749 and SNI 58928.
The numerical analysis was partially carried at GuaCAL (Guanajuato Computational Astroparticle Lab).

\appendix

\section{6-dimensional field decomposition}
\label{app:6decomp}

In this appendix we show the decomposition of the 6-dimensional fields in the model into 4-dimensional ones and describe which representations have zero modes.
There are three sets of bulk fields: the gauge 6-dimensional vector fields denoted by $\textbf{A}_M$ (with $M=0,1,2,3,5,6$);
three 6-dimensional chiral left fermions ($\textbf{u}^c_i$, $i=1,2,3$); and the three scalar fields that make up the flavour triplet $\sigma$. 

The orbifold $\mathbb{Z}_2$ projection, 
$
z\sim -z,$
acts on every field $\Phi$, denoted generically, as
\begin{equation}
\Phi(x,z)=P \mathcal{P}_{56} \Phi(x,-z),
\end{equation}
where $P$ is the symmetry transformation constrained by Eq.(\ref{eq:Pcon}) and $\mathcal{P}_{56}$ is the fifth and sixth parity operator for the corresponding Poincar\'e representation.
Each 6-dimensional field is decomposed into an infinite tower of 4-dimensional fields.
Only the 4-dimensional fields that have a positive eigenvalue of $P \mathcal{P}_{56}$ can have zero modes, and these are the only ones relevant at low energies.

The gauge vector fields must fulfill the orbifold projection
\begin{equation}
\textbf{A}_M(x,z)=P \mathcal{P}_{56} \textbf{A}_M(x,-z) = \begin{cases} A_\mu(x,z)=A_\mu(x,-z),\\
  A_{5,6}(x,z)=-A_{5,6}(x,-z), \end{cases}
\end{equation}
where $P=1$ for the gauge fields, as these are neutral under family transformations. One can see that, in contrast to the 4-dimensional vectors,  the extra components do not have zero modes.
Therefore the gauge symmetry is preserved and consistent with the orbifold.

The 6-dimensional chiral left fermion $\textbf{u}^c_i$ are family singlets, and therefore their $P$ action is trivial.
The parity transformation for fermions involves the $\Gamma_M$ matrices from the 6-dimensional Clifford algebra satisfying \cite{Scrucca:2003ut,Hosotani:2004ka}
\begin{equation}
\{\Gamma_M,\Gamma_N\}=2\eta_{MN},
\end{equation}
from which one can build the 6-dimensional chiral operator $\Gamma_7=\Gamma_0\Gamma_1\Gamma_2\Gamma_3\Gamma_5\Gamma_6 $.
The left 6-dimensional chiral fermions are defined by $\Gamma_7 \textbf{u}^c_i=-\textbf{u}^c_i$.

The orbifold projection acts on the bulk quarks as
\begin{equation}
\begin{split}
\textbf{u}^c_i(x,z)=&\Gamma_5\Gamma_6 \textbf{u}^c_i(x,-z)\\
=&(\Gamma_0\Gamma_1\Gamma_2\Gamma_3)(\Gamma_0\Gamma_1\Gamma_2\Gamma_3\Gamma_5\Gamma_6 )\textbf{u}^c_i(x,-z)\\
=&(\Gamma_0\Gamma_1\Gamma_2\Gamma_3)\Gamma_7\textbf{u}^c_i(x,-z)\\
=&-(\Gamma_0\Gamma_1\Gamma_2\Gamma_3)\textbf{u}^c_i(x,-z).
\end{split}
\end{equation}
It is now useful to separate the 6-dim chiral fermion $\textbf{u}^c_i=(u^c_{Li},\overline{u}^c_{Ri})^T$ into a left and a right 4-dim chiral fermion pair so that 
\begin{equation}
\begin{split}
\textbf{u}^c_i(x,z)
=&-(\Gamma_0\Gamma_1\Gamma_2\Gamma_3)(u^c_{Li},\overline{u}^c_{Ri})^T\\
=&-(\gamma^5 u^c_{Li},\gamma^5\overline{u}^c_{Ri})^T\\
=&(u^c_{Li},-\overline{u}^c_{Ri})^T,
\end{split}
\end{equation}
One sees that only the 4-dim left chiral fermion $u^c_{Li}$ (denoted simply by $u^c_{i}$) has zero modes, as desired.

Finally the scalars, which are neutral under all Poincar\'e transformations, transform trivially under the parity transformation but not under the family transformation $P$.
This allows only one of the three scalars to have a zero mode, implying the VEV alignment described in Eq(\ref{eq:bcOfSigmaVEV}).

\section{$A_4$ basis}
\label{app:a4}
The $A_4$ group can be defined by the presentation
\begin{equation}
A_4\simeq \{S,T|S^2=T^3=(ST)^2=1\}.
\end{equation}
It has 4 irreducible representations, which in the Ma-Rajasekaran basis transform as
\begin{equation}\begin{split}
\textbf{1}&:\ \ \ \ S=1,\ \ \ \ T=1,\\
\textbf{1}'&:\ \ \ \ S=1,\ \ \ \ T=\omega,\\
\textbf{1}''&:\ \ \ \ S=1,\ \ \ \ T=\omega^2,\\
\textbf{3}&:\ \ \ \ S=\left(\begin{array}{ccc} 1&0&0\\ 0&-1&0\\ 0&0&-1
\end{array}\right),\ \ \ \ T=\left(\begin{array}{ccc} 0&1&0\\ 0&0&1\\ 1&0&0
\end{array}\right),\\
\end{split}\end{equation}
where $\omega=e^{2i\pi/3}$. These define the invariant non trivial contractions as
\begin{equation}\begin{split}
\textbf{1}'\otimes\textbf{1}'=&\textbf{1}'',\ \ \ \textbf{1}''\otimes\textbf{1}''=\textbf{1}',\ \ \ \textbf{1}''\otimes\textbf{1}'=\textbf{1},\\
\textbf{3}&\otimes \textbf{3} =\textbf{1}+\textbf{1}'+\textbf{1}''+\textbf{3}_1+\textbf{3}_2.
\end{split}\end{equation}
The contractions of two triplets $\textbf{3}_a\sim (a_1,a_2,a_3)$ and $\textbf{3}_b\sim (b_1,b_2,b_3)$ are decomposed as
\begin{equation}\begin{split}
\textbf{3}_a\otimes \textbf{3}_b\to \textbf{1}&=a_1b_1+a_2b_2+a_3b_3,\\\textbf{3}_a\otimes \textbf{3}_b\to \textbf{1}'&=a_1b_1+\omega^2 a_2b_2+\omega a_3b_3,\\
\textbf{3}_a\otimes \textbf{3}_b\to \textbf{1}''&=a_1b_1+\omega a_2b_2+\omega^2 a_3b_3,\\
\textbf{3}_a\otimes \textbf{3}_b\to \textbf{3}_1&=(a_2b_3,a_3b_1,a_1b_2),\\
\textbf{3}_a\otimes \textbf{3}_b\to \textbf{3}_2&=(a_3b_2,a_1b_3,a_2b_1).\\
\end{split}\end{equation}
The two invariant contractions of three triplets in the Ma-Rajasenkaran basis are
\begin{equation}\begin{split}
\textbf{3}_a\otimes \textbf{3}_b  \otimes \textbf{3}_ c \to \textbf{1}_1 = a_1 b_2 c_3 + a_2 b_3 c_1+ a_3 b_1 c_2 ,\\
\textbf{3}_a\otimes \textbf{3}_b  \otimes \textbf{3}_ c \to \textbf{1}_2 = a_1 b_3 c_2 + a_2 b_1 c_3+ a_3 b_2 c_1 .\\
\end{split}\end{equation}

\section{Boundary Condition Matrix $P$}
\label{app:Pmatrix}
The most general matrix $P$ that satisfies the conditions listed in Eq. (\ref{eq:bcOfSigmaVEV}) in terms of the three independent parameters  ($\epsilon^\sigma_{1,2}, \varphi$) of the VEV alignment of $\sigma$ in Eq. (\ref{eq:SigmaVEVBoundary}) is given by

\begin{equation}
P\left(\epsilon^{\sigma}_{1},\epsilon^{\sigma}_{2},\varphi\right) = \left(1+ (\epsilon^{\sigma}_{1})^2 + (\epsilon^{\sigma}_{2})^2 \right)^{-1}  \left(\begin{array}{ccc} -1+ (\epsilon^{\sigma}_{1})^2 - (\epsilon^{\sigma}_{2})^2  & 2 \epsilon^{\sigma}_{1}  \epsilon^{\sigma}_{2} e^{i\varphi} &  2 \epsilon^{\sigma}_{1} e^{i\varphi} \\
 2 \epsilon^{\sigma}_{1}  \epsilon^{\sigma}_{2} e^{-i\varphi} & -1-(\epsilon^{\sigma}_{1})^2 +(\epsilon^{\sigma}_{2})^2  &  2 \epsilon^{\sigma}_{2} \\
 2 \epsilon^{\sigma}_{1} e^{-i\varphi} &   2 \epsilon^{\sigma}_{2} & 1-(\epsilon^{\sigma}_{1})^2 -(\epsilon^{\sigma}_{2})^2  \end{array}\right).
\label{eq:Pmatrix}
\end{equation}

\bibliographystyle{utphys}
\bibliography{bibliography}
\end{document}